\documentclass[%
 reprint,
 amsmath,amssymb,
 aps,
]{revtex4-2}

\usepackage{breqn}
\usepackage{amsmath}
\usepackage{color}
\usepackage{graphicx}
\usepackage{dcolumn}
\usepackage{bm}
\usepackage{hyperref}

\hypersetup{
    colorlinks=true,
    linkcolor=blue,
    citecolor=blue,
    filecolor=magenta,      
    urlcolor=blue,
    pdftitle={Overleaf Example},
    pdfpagemode=FullScreen,
    pdftitle={A Fully Analytic Expression for the N-Body 2PN Hamiltonian},
    pdfauthor={Felix M. Heinze, Gerhard Schäfer, and Bernd Brügmann}
    }

\begin{document}

\preprint{APS/123-QED}

\title{A Fully Analytic Expression for the $N$-Body 2PN Hamiltonian}

\author{Felix M. Heinze}
\author{Gerhard Schäfer}
\author{Bernd Brügmann}
\affiliation{Friedrich-Schiller-Universität Jena$,$ Theoretisch-Physikalisches Institut$,$ 07743 Jena$,$ Germany}
\date{\today}

\begin{abstract}
We present a complete analytic expression for the $N$-body Hamiltonian of nonspinning point particles at second post-Newtonian (2PN) order in the Arnowitt-Deser-Misner (ADM) gauge. Recent work reduced the general $N$-body Hamiltonian to a form containing a single unresolved integral term, which prevented a fully explicit closed-form representation. Here, we evaluate this integral analytically and thereby obtain a fully explicit closed-form expression for the general $N$-body 2PN Hamiltonian. We outline the main steps of the derivation and validate the analytic expression through direct numerical comparisons with the original integral representation.
\end{abstract}

                              
\maketitle


\section{Introduction}
\label{sec:introduction}

The post-Newtonian (PN) approximation provides a systematic framework for describing gravitational dynamics in general relativity when characteristic velocities are small compared with the speed of light and the inter-body gravitational fields are weak. In this regime, Einstein's equations can be solved perturbatively as an asymptotic series expansion in powers of the small parameter $\epsilon \sim v^2/c^2 \sim GM/(rc^2)$ (where the last relation comes from the virial theorem). The highest power of $\epsilon$ retained in the series expansion defines the order of the post-Newtonian approximation \cite{Blanchet2024}. Its applicability to compact objects does not require their internal gravitational fields to be weak: sufficiently well-separated bodies can be described by an effective point-particle model to account for their orbital dynamics \cite{FutamaseItoh2007}. When dimensional regularization is applied, such point-particle models of compact binaries are consistent through 4PN order~\cite{SchaeferJaranowski2024}.

For systems containing several gravitating bodies, PN methods provide a practical means of retaining relativistic effects in numerical studies of orbital evolution without the need to solve the full Einstein equations. Applications of the PN approximation range from high-precision modeling of solar-system dynamics~\cite{ParkEtAl2021} and the interpretation of binary-pulsar timing measurements~\cite{WeisbergHuang2016,KramerEtAl2021} to the evolution of relativistic multiple systems~\cite{KupiEtAl2006,NaozEtAl2013} and the modeling of the dynamics and gravitational-wave emission of inspiraling compact binaries~\cite{BlanchetDamourIyer1995}. In addition, several modern $N$-body codes for modeling dense stellar systems allow the inclusion of PN corrections up to 3.5PN order for compact-object interactions~\cite{RantalaEtAl2023,SpurzemKamlah2023}. However, a fully explicit analytic description of general $N$-body dynamics at 2PN order has remained incomplete until the present work, while the corresponding 3PN description is still incomplete for general $N$-body systems. A consistent treatment requires interactions involving couplings among several bodies that arise from the nonlinearity of Einstein's equations. Such interactions generally cannot be recovered by adding isolated two-body PN corrections to Newtonian $N$-body dynamics. The importance of such cross-terms for secular evolution has been demonstrated explicitly for hierarchical triples and systems containing a dominant central mass~\cite{Will2014}.

The broad range of astrophysical applications has driven major advances in PN methods and the development of complementary formalisms \cite{BlanchetDamour1986, PoujadeBlanchet2002, PatiWill2000, FutamaseItoh2007, Levi:2018nxp}. The canonical formulation of Arnowitt, Deser, and Misner (ADM) provides a natural framework for conservative dynamics \cite{ADM1962, SchaeferJaranowski2024}. Eliminating the gravitational field degrees of freedom to the required PN order yields a reduced Hamiltonian in the particle positions and canonical momenta. Hamilton's equations give the equations of motion, and the canonical structure facilitates the analysis of conserved quantities and the design of numerical integration schemes.

The general $N$-body equations of motion at 1PN order were already derived in 1938 by Einstein, Infeld, and Hoffmann \cite{EIH1938}. At 2PN order, formal expressions for the $N$-body gravitational interaction were obtained by Ohta et al.~\cite{OhtaEtAl1974b,OhtaEtAl1974a}, but their evaluation for arbitrary particle numbers was obstructed by unevaluated spatial integrals. Damour and Sch\"afer \cite{DamourSchaefer1985} subsequently clarified and corrected computational issues in the results for two-body systems and established a relation between 2PN Lagrangians in different coordinate systems. Sch\"afer \cite{Schaefer1987} extended the explicit ADM Hamiltonian at 2PN order to three-body systems by evaluating the required three-body contribution to the static potential associated with the transverse-traceless gravitational field. Subsequent work corrected the three-body reduction of static $G^3$ potential terms in the three-body Hamiltonian and used the results in numerical studies of relativistic three-body motion \cite{LoustoNakano2008,GalavizBruegmann2011}. An independent treatment by Chu \cite{Chu2009}, based on perturbative field theory, obtained the general $N$-body effective Lagrangian through 2PN order up to unevaluated integrals, leaving the explicit four-body interaction as an open problem. More recently, Heinze, Schäfer, and Brügmann \cite{HeinzeSchaeferBruegmann2026} reduced the general ADM 2PN Hamiltonian to an analytic expression containing a single remaining unevaluated integral, denoted by $I^{\ln}_{ab;cd}$, and developed methods for its accurate numerical evaluation and for integrating the resulting equations of motion. In a complementary harmonic-coordinate treatment, Huang, Yang, and Ni \cite{HuangYangNi2026} derived a computable formulation of the general point-mass $N$-body equations of motion at 2PN order. Their formulation separates the equations into closed analytic terms and a regularized, numerically evaluable non-closed-form integral contribution. These developments made the general 2PN dynamics computationally accessible in both the ADM and harmonic formulations, while a fully explicit closed-form expression remained unavailable.

In this paper, we evaluate the remaining integral $I^{\ln}_{ab;cd}$ analytically and thereby complete the closed-form ADM Hamiltonian for an arbitrary number of point particles up to 2PN order. We outline the main steps of the derivation and provide more details in the Supplemental Material~\cite{SupplementalMaterial}. Numerical comparisons with the original integral representation are used for validation.

The paper is structured as follows. In Sec.~\ref{sec:pn_formalism} we summarize the results for the general $N$-body post-Newtonian Hamiltonian for nonspinning point particles up to 2PN order, and provide an analytic expression for the missing integral term $I^{\mathrm{ln}}_{ab;cd}$. In Sec.~\ref{sec:numerical_validation} we present the numerical validation of our main results and Sec.~\ref{sec:conclusion} concludes with some final remarks. Throughout this paper we use geometric units where $G=c=1$.

\section{The 2PN N-Body post-Newtonian Hamiltonian}
\label{sec:pn_formalism}
In this section we present the general $N$-body post-Newtonian Hamiltonian for nonspinning point particles up to 2PN order in the ADM gauge. The presentation largely follows Sec.~II of our previous work \cite{HeinzeSchaeferBruegmann2026}, supplemented by the analytic evaluation of the remaining integral $I^{\mathrm{ln}}_{ab;cd}$. The Hamiltonian can be written as 
\begin{equation}
    H = H_{\mathrm{N}} + H_{\mathrm{1PN}} + H_{\mathrm{2PN}},
\end{equation}
where the subscript N describes the Newtonian contribution, 1PN the first-order post-Newtonian correction, and 2PN the second-order post-Newtonian correction. The dynamics described by this Hamiltonian are fully conservative. Dissipative effects due to the emission of gravitational radiation first enter at the 2.5PN order. Adding the leading-order 2.5PN and next-to-leading-order 3.5PN radiation-reaction terms to the Hamiltonian can be done for general $N$-body systems \cite{JaranowskiSchaefer1997}, but they are not considered here. 
\begin{widetext}
\noindent
In the Newtonian theory, one obtains
\begin{equation}
    H_{\mathrm{N}} = \frac{1}{2} \sum_a \frac{p_a^2}{m_a} - \frac{1}{2} \sum_a \sum_{b \neq a} \frac{m_a m_b}{r_{ab}},
\end{equation}
and the first-order post-Newtonian correction to the Hamiltonian is given by
\begin{equation}
\begin{aligned}
H_{\mathrm{1 P N}}= &-\frac{1}{8} \sum_a m_a \frac{p_a^4}{m_a^4} -\frac{1}{4} \sum_a \sum_{b \neq a} \frac{m_a m_b}{r_{a b}}\Bigg\{6 \frac{p_a^2}{m_a^2} -7 \frac{\mathbf{p}_a \cdot \mathbf{p}_b}{m_a m_b} -\frac{\left(\mathbf{n}_{a b} \cdot \mathbf{p}_a\right)\left(\mathbf{n}_{a b} \cdot \mathbf{p}_b\right)}{m_a m_b}\Bigg\} +\frac{1}{2} \sum_a \sum_{b \neq a} \sum_{c \neq a}  \frac{m_a m_b m_c}{r_{a b} r_{a c}},
\end{aligned}
\end{equation}
where $\mathbf{x}_a$ and $\mathbf{p}_a$ are the position and canonical momentum, $r_{ab} = |\mathbf{x}_a-\mathbf{x}_b|$, $\mathbf{n}_{ab}=(\mathbf{x}_a-\mathbf{x}_b)/r_{ab}$, and $m_a$ is the mass of the $a$th body. The $N$-body Hamiltonian at the second PN order contains the static potential term
\begin{equation}
    U^{\mathrm{TT}} = - \frac{1}{4 \pi} \int d^3x \sum_{i,j,k} \partial_k f_{ij}^{\mathrm{TT}} \ \partial_k f_{ij}^{\mathrm{TT}},
\end{equation}
with
\begin{multline}
    f_{ij}^{\mathrm{TT}} = f_{ij} - \frac{1}{8} \sum_a \sum_{b \neq a} m_a m_b \Bigg[ \delta_{ij} \Bigg( \frac{1}{r_a r_b} - \frac{2}{r_a r_{ab}} + \frac{2}{r_{ab}^2} \mathbf{n}_a \cdot \mathbf{n}_{ab}\Bigg) \\
    + \frac{\partial}{\partial x^i} \frac{\partial}{\partial x^j} \Bigg( \ln(r_a + r_b + r_{ab}) - \frac{r_a}{r_{ab}} + \frac{r_a^2}{2r_{ab}^2} \mathbf{n}_a \cdot \mathbf{n}_{ab}\Bigg) - \frac{4}{r_{ab}^2} (n_a^i n_{ab}^j + n_a^j n_{ab}^i) \Bigg]
\end{multline}
and
\begin{equation}
    f_{ij} = \sum_a \sum_{b \neq a} m_a m_b \frac{\partial}{\partial x_a^i} \frac{\partial}{\partial x_b^j} \ln(r_a + r_b + r_{ab}),
\end{equation}
where $r_a = |\mathbf{x}-\mathbf{x}_a|$, $\mathbf{n}_a=(\mathbf{x}-\mathbf{x}_a)/r_a$, and $\mathrm{TT}$ indicates the transverse-traceless part of the gravitational potential tensor $f_{ij}$. Due to the double sum in $f_{ij}^{\mathrm{TT}}$, the potential contains
contributions involving up to four distinct particle labels. We organize
these contributions as
\begin{equation}
    U^{\mathrm{TT}} = U^{\mathrm{TT}}_{(2)} + U^{\mathrm{TT}}_{(3)} + U^{\mathrm{TT}}_{(4)},
\end{equation}
where $U^{\mathrm{TT}}_{(2)}$ is the two-point correlation function, for which only two particle indices in the sums are distinct, $U^{\mathrm{TT}}_{(3)}$ is the three-point correlation function, for which three particle indices are distinct, and $U^{\mathrm{TT}}_{(4)}$ is the four-point correlation function, for which all four particle indices are distinct. The two-point correlation function has been calculated in \cite{OhtaEtAl1974a, DamourSchaefer1985} for two particles, and for $N$ particles it is given by
\begin{equation}
    U^{\mathrm{TT}}_{(2)} = - \frac{1}{4} \sum_a \sum_{b \neq a} \frac{m_a^2m_b^2}{r_{ab}^3}.
\end{equation}
The three-point correlation function has been calculated by Schäfer \cite{Schaefer1987} and is given by
\begin{equation}
\begin{aligned}
    U^{\mathrm{TT}}_{(3)} = - \frac{1}{64} \sum_a \sum_{b \neq a} \sum_{c \neq a,b} \frac{m_a^2 m_b m_c}{r_{ab}^3r_{ac}^3r_{bc}} \Big[& 18r_{ab}^2r_{ac}^2 - 60r_{ab}^2r_{bc}^2 - 24r_{ab}^2r_{ac}(r_{ab}+r_{bc}) +60 r_{ab}r_{ac}r_{bc}^2 \\
    &+56 r_{ab}^3r_{bc} - 72 r_{ab}r_{bc}^3 + 35 r_{bc}^4 + 6 r_{ab}^4  \Big].
    \label{eq:utt3}
\end{aligned}
\end{equation}
For $N=2$, only $U^{\mathrm{TT}}_{(2)}$ contributes, $U^{\mathrm{TT}}_{(3)}$ must additionally be included for $N=3$, and systems with $N\geq4$ require all three contributions, including $U^{\mathrm{TT}}_{(4)}$. No higher-point correlation function occurs at 2PN order. Heinze, Schäfer, and Brügmann \cite{HeinzeSchaeferBruegmann2026} derived an expression for the four-point correlation function, given by
\begin{equation}
\begin{aligned}
    U^{\mathrm{TT}}_{(4)} = &- \frac{1}{64} \sum_a \sum_{b \neq a} \sum_{c \neq a,b} \sum_{d \neq a,b,c} \frac{m_a m_b m_c m_d}{r_{ab}^3 r_{cd}^3 r_{ad}^3 r_{bc}^3} \Bigg\{ 16 \frac{r_{ab}^3 r_{bc}^3  r_{cd}^2 r_{ad}^2}{r_{bd}} - 30 r_{ad}^4 r_{bc}^3 (r_{ad}^2 + r_{bc}^2 - r_{ac}^2 - r_{bd}^2) \\
    &- 24 r_{bc}^3 r_{ab}^2 r_{cd}^2 r_{ad}^2 + r_{ab}^2 (r_{bd}^2-r_{bc}^2-r_{cd}^2) \Bigg(16 \frac{r_{ab} r_{ad}^3 r_{bc}^2}{r_{ac}+r_{bc}+r_{ab}} - 8 r_{ad}^3 r_{bc}^2 + r_{ab} r_{cd}^2 (r_{ac}^2-r_{ad}^2-r_{cd}^2) \Bigg) \Bigg\} \\
    &+ \frac{1}{4 \pi} \sum_a \sum_{b \neq a} \sum_{c \neq a,b} \sum_{d \neq a,b,c} m_a m_b m_c m_d \; I^{\mathrm{ln}}_{ab;cd}.
    \label{eq:UTT4}
\end{aligned}
\end{equation}
This expression contains the integral term
\begin{equation}
\begin{aligned}
    I^{\mathrm{ln}}_{ab;cd} &= \sum_{i,j} \frac{\partial^2}{\partial x_a^i \partial x_b^j} \frac{\partial^2}{\partial x_c^i \partial x_d^j}  \int d^3x \frac{\ln(r_a+r_b+r_{ab})}{r_c r_d},
    \label{eq:ln_integral}
\end{aligned}
\end{equation}
which obstructs a fully explicit closed-form representation of the $N$-body 2PN Hamiltonian. Eq.~\eqref{eq:ln_integral} is understood with at least two of the particle derivatives applied to the integrand before the spatial integration. For four distinct, noncoincident particle positions, the resulting integrand is locally integrable at every particle position and decays sufficiently rapidly at spatial infinity so that no regularization is required. This is the integral which was evaluated numerically in \cite{HeinzeSchaeferBruegmann2026} to enable the computation of the complete Hamiltonian.
\\
\\
We found a fully analytic solution to the integral expression in Eq.~\eqref{eq:ln_integral}, which is given by 
\begin{equation}
    \begin{aligned}
I_{a b ; c d}^{\mathrm{ln}}&=\frac{\pi}{r_{a b}^3 r_{c d}^3}\Bigg\{\left(r_{a c}+r_{a d}+r_{b c}+r_{b d}\right)\left(2 r_{a c} r_{b d}+2 r_{a d} r_{b c}-r_{a b} r_{c d}\right) - \left(r_{a d}+r_{b c}-r_{a c}-r_{b d}\right)\left(r_{a d}^2+r_{b c}^2-r_{a c}^2-r_{b d}^2\right)  \\
&+ 2 D_{\{a,b,c,d\}} {\Bigg[\frac{r_{a b}^2-\left(r_{a d}+r_{b d}\right)\left(r_{a c}+r_{b c}\right)}{F_{ab;cd}}+\frac{r_{c d}^2-\left(r_{a c}+r_{a d}\right)\left(r_{b c}+r_{b d}\right)}{F_{cd;ab}}} +\frac{r_{a b} r_{c d}\left(r_{a b} r_{c d}+r_{a d} r_{b c}\right)}{r_{a d} r_{b c}\left(r_{a b} r_{c d}+r_{a c} r_{b d}+r_{a d} r_{b c}\right)}\Bigg] \\
& -r_{a b} r_{c d}\left[2\left(r_{a b}+r_{c d}\right)+\frac{r_{a b} r_{c d}\left(r_{a b}+r_{c d}+2 r_{a d}+2 r_{b c}+r_{a c}+r_{b d}\right)}{r_{a d} r_{b c}}\right]-4 \Delta_{\{a,b,c,d\}}  \Phi_{\{a,b,c,d\}} \Bigg\},
    \label{eq:I_ln_solution}
\end{aligned}
\end{equation}
where 
\begin{equation}
    F_{ab;cd} = (r_{ab}+r_{ac}+r_{bc})(r_{ab}+r_{ad}+r_{bd})
\end{equation}
is the product of the perimeters of the two triangular faces sharing the edge $(ab)$,
\begin{equation}
    D_{\{a,b,c,d\}} = \frac{1}{2}\Big[r_{a b} F_{cd;ab}+r_{c d} F_{ab;cd} +\left(r_{a c}+r_{a d}+r_{b c}+r_{b d}\right)\left(r_{a c} r_{b d}+r_{a d} r_{b c}-r_{a b} r_{c d}\right)\Big]
\end{equation}
is a positive, symmetric cubic invariant of the inter-particle distances,
\begin{equation}
\Delta_{\{a,b,c,d\}} = \left|
(\mathbf{x}_a-\mathbf{x}_b)\cdot
\bigl[(\mathbf{x}_a-\mathbf{x}_c)\times
      (\mathbf{x}_a-\mathbf{x}_d)\bigr]
\right| = \sqrt{\frac{1}{2}\left(r_{a b} r_{c d}+r_{a c} r_{b d}+r_{a d} r_{b c}\right) F_{ab;cd} F_{cd;ab}-D_{\{a,b,c,d\}}^2}
\end{equation}
is six times the volume of the tetrahedron that has the four particles $\{a,b,c,d\}$ at its vertices, and
\begin{equation}
    \Phi_{\{a,b,c,d\}} = \arctan \left( \frac{\Delta_{\{a,b,c,d\}}}{D_{\{a,b,c,d\}}} \right)
\end{equation}
is a dimensionless angle that depends only on the shape of the particle configuration and is unchanged under uniform rescaling. The quantities $D_{\{a,b,c,d\}}$, $\Delta_{\{a,b,c,d\}}$ and $\Phi_{\{a,b,c,d\}}$ depend only on the particle labels $\{a,b,c,d\}$ but not on their ordering. They are therefore fully symmetric under the exchange of any particle-label pair.
\\
\\
The derivation of Eq.~\eqref{eq:I_ln_solution} was developed with substantive assistance from OpenAI's large language model \textsc{GPT-6 Astra}. The authors supplied the mathematical input, including analytic derivations for related integrals appearing in $U^{\mathrm{TT}}_{(4)}$ and several approaches to solving $I^{\ln}_{ab;cd}$. The derivation was independently verified by the  authors and the final expression was additionally validated against direct numerical evaluations of the original integral representation (see Sec. \ref{sec:numerical_validation}), which in this context plays the important role of a fully independent consistency check. 
\\
\\
The first step of the derivation is the reduction of the 3D integral over $\mathbb{R}^3$ to a bounded 2D parameter integral by using the identity~\cite{JaranowskiSchafer1998} $\nabla^2 \ln(r_a+r_b+r_{ab}) = 1/(r_a r_b)$ and then Fourier-transforming the pair products using Feynman parameters. The spatial and Fourier integrations can then be performed analytically, which results in
\begin{equation}
    I_{a b ; c d}^{\ln }=- (\boldsymbol{\nabla}_a \cdot \boldsymbol{\nabla}_c) (\boldsymbol{\nabla}_b \cdot \boldsymbol{\nabla}_d) \int_0^\pi d \theta \int_0^\pi d \phi \, K(M, S),
    \label{eq:2d_parameter_integral}
\end{equation}
with
\begin{equation}
    K(M, S)=\frac{M^2-S}{\sqrt{S}} \arctan \left( \frac{\sqrt{S}}{M} \right)+M \ln \left( \frac{M^2+S}{\ell^2} \right),
\end{equation}
\begin{equation}
\begin{gathered}
u=\frac{1+\cos \theta}{2}, \quad v=\frac{1+\cos \phi}{2}, \quad m=r_{a b} \sqrt{u(1-u)}, \quad n=r_{c d} \sqrt{v(1-v)},
\end{gathered}
\end{equation}
\begin{equation}
    \begin{gathered}
M=m+n, \quad S = u v r_{a c}^2+u(1-v) r_{a d}^2+(1-u) v r_{b c}^2 +(1-u)(1-v) r_{b d}^2-m^2-n^2.
\end{gathered}
\end{equation}
Here \(\ell\) is an arbitrary fixed reference length. Its contribution vanishes after differentiation and the apparent singularity at \(S=0\) is removable. This expression already allows for a much more efficient numerical evaluation using two-dimensional quadrature.

To obtain a complete closed-form expression, one can evaluate the contracted particle derivatives and use integration by parts in parameter space to separate boundary contributions from the remaining scalar integrals. The boundary contributions can be evaluated in elementary functions. For the remaining terms, we introduce the perpendicular separation $H$ between the lines through the pairs (a,b) and (c,d), keeping their directions and longitudinal offsets fixed. A divergence identity isolates the angular contribution, while differentiation with respect to $H^2$, combined with substitutions that rationalize the parameter square roots, reduces the remaining integrations to a finite set of elementary moments. The angular contribution is fixed by a first-derivative identity, while the rational contribution is established by matching second derivatives with respect to $H^2$. The integration constants are fixed by the behavior as $H\to\infty$.

Finally, algebraic identities combine the boundary and interior contributions, cancel the intermediate logarithms, and express the result entirely through the six inter-particle distances and elementary functions of those distances. The resulting expression was further simplified via symbolic manipulations. The sum over distinct $a$, $b$, $c$ and $d$ allows for further simplifications. For the remaining summed contribution in $U^{\mathrm{TT}}_{(4)}$ we obtain
\begin{equation}
\begin{aligned}
\frac{1}{4 \pi}& \sum_a \sum_{b \neq a} \sum_{c \neq a,b} \sum_{d \neq a,b,c}  m_a m_b m_c m_d \, I_{a b ; c d}^{\ln } = \sum_a \sum_{b \neq a} \sum_{c \neq a,b} \sum_{d \neq a,b,c}
\frac{m_a m_b m_c m_d}{r_{ab}^{3}r_{cd}^{3}}
\Bigg[
\frac{
2r_{bc}r_{cd}\bigl(r_{ad}^{2}-r_{ab}^{2}-r_{bc}r_{bd}\bigr)
}{
r_{bc}+r_{bd}+r_{cd}
} \\
& +\frac{r_{ab}^{2}r_{cd}}{6}
\left(
\frac{6r_{ad}r_{bd}}
{r_{ab}r_{cd}+r_{ad}r_{bc}+r_{ac}r_{bd}}
+
\frac{2r_{cd}^{2}+3r_{ac}(r_{ad}-r_{cd})}
{r_{ac}r_{ad}}
\right)
-r_{ad}^{2}(r_{ad}+r_{bc}-2r_{ac}) 
-\Delta_{\{a,b,c,d\}}\Phi_{\{a,b,c,d\}}
\Bigg].
\end{aligned}
\label{eq:H_ln_compact}
\end{equation}
With this, the four-point correlation function can be written as
\begin{equation}
\begin{aligned}
U^{\mathrm{TT}}_{(4)}
={}&-\frac{1}{64} \sum_a \sum_{b \neq a} \sum_{c \neq a,b} \sum_{d \neq a,b,c}
\frac{m_a m_b m_c m_d}{r_{ab}^{3}r_{cd}^{3} r_{ad}}
\Bigg\{
35r_{ad}^{4}+34r_{ad}^{2}r_{bc}^{2}+10r_{ab}^{2}r_{ad}^{2}+r_{ac}^{2}r_{bd}^{2}-70r_{ad}^{2}r_{bd}^{2}-10r_{bd}^{2}r_{cd}^{2} \\
&+17r_{ab}^{2}r_{cd}^{2} - 32 r_{ab}^2 r_{ad} r_{cd}
+16r_{cd}\Bigg(
\frac{
(r_{cd}^{2}-8r_{ac}r_{ad})
(r_{bd}^{2}-r_{ab}^{2}-r_{ac}r_{ad})
}{
r_{ac}+r_{ad}+r_{cd}
} - \frac{
r_{ab}^2r_{ac}(r_{ab}r_{cd}+4r_{ad}r_{bc})
}{
r_{ab}r_{cd}+r_{ac}r_{bd}+r_{ad}r_{bc}
}
\Bigg)\Bigg\} \\
&- \sum_a \sum_{b \neq a} \sum_{c \neq a,b} \sum_{d \neq a,b,c}
\frac{m_a m_b m_c m_d}{r_{ab}^{3}r_{cd}^{3}}\Delta_{\{a,b,c,d\}}\Phi_{\{a,b,c,d\}},
\end{aligned}
\label{eq:UTT4_compact}
\end{equation}
and combined with previous results, the 2PN part of the $N$-body Hamiltonian can finally be written as
\begin{equation}
\begin{aligned}
     H_{\mathrm{2PN}}= & \frac{1}{16} \sum_a m_a\left(\frac{p_a^2}{m_a^2}\right)^3+\frac{1}{16} \sum_{a} \sum_{b \neq a} \frac{m_a m_b}{r_{a b}}\left\{10\left(\frac{p_a^2}{m_a^2}\right)^2-11 \frac{p_a^2 p_b^2}{m_a^2 m_b^2}-2 \frac{\left(\mathbf{p}_a \cdot \mathbf{p}_b\right)^2}{m_a^2 m_b^2}\right. \\
    & \left.+10 \frac{p_a^2\left(\mathbf{n}_{a b} \cdot \mathbf{p}_b\right)^2}{m_a^2 m_b^2}-12 \frac{\left(\mathbf{p}_a \cdot \mathbf{p}_b\right)\left(\mathbf{n}_{a b} \cdot \mathbf{p}_a\right)\left(\mathbf{n}_{a b} \cdot \mathbf{p}_b\right)}{m_a^2 m_b^2}-3 \frac{\left(\mathbf{n}_{a b} \cdot \mathbf{p}_a\right)^2\left(\mathbf{n}_{a b} \cdot \mathbf{p}_b\right)^2}{m_a^2 m_b^2}\right\} \\
    & +\frac{1}{8} \sum_{a} \sum_{b \neq a} \sum_{c \neq a} \frac{m_a m_b m_c}{r_{a b} r_{a c}}\Bigg\{18 \frac{p_a^2}{m_a^2}+14 \frac{p_b^2}{m_b^2}-2 \frac{\left(\mathbf{n}_{a b} \cdot \mathbf{p}_b\right)^2}{m_b^2}-50 \frac{\mathbf{p}_a \cdot \mathbf{p}_b}{m_a m_b}+17 \frac{\mathbf{p}_b \cdot \mathbf{p}_c}{m_b m_c} \\
    & -14 \frac{\left(\mathbf{n}_{a b} \cdot \mathbf{p}_a\right)\left(\mathbf{n}_{a b} \cdot \mathbf{p}_b\right)}{m_a m_b}+14 \frac{\left(\mathbf{n}_{a b} \cdot \mathbf{p}_b\right)\left(\mathbf{n}_{a b} \cdot \mathbf{p}_c\right)}{m_b m_c}+\mathbf{n}_{a b} \cdot \mathbf{n}_{a c} \frac{\left(\mathbf{n}_{a b} \cdot \mathbf{p}_b\right)\left(\mathbf{n}_{a c} \cdot \mathbf{p}_c\right)}{m_b m_c}\Bigg\} \\
    & +\frac{1}{8} \sum_{a} \sum_{b \neq a} \sum_{c \neq a} \frac{m_a m_b m_c}{r_{a b}^2}\Bigg\{2 \frac{\left(\mathbf{n}_{a b} \cdot \mathbf{p}_a\right)\left(\mathbf{n}_{a c} \cdot \mathbf{p}_c\right)}{m_a m_c}+2 \frac{\left(\mathbf{n}_{a b} \cdot \mathbf{p}_b\right)\left(\mathbf{n}_{a c} \cdot \mathbf{p}_c\right)}{m_b m_c} +5 \mathbf{n}_{a b} \cdot \mathbf{n}_{a c} \frac{p_c^2}{m_c^2}\\
    &-\mathbf{n}_{a b} \cdot \mathbf{n}_{a c} \frac{\left(\mathbf{n}_{a c} \cdot \mathbf{p}_c\right)^2}{m_c^2}-14 \frac{\left(\mathbf{n}_{a b} \cdot \mathbf{p}_c\right)\left(\mathbf{n}_{a c} \cdot \mathbf{p}_c\right)}{m_c^2}\Bigg\} +\frac{1}{4} \sum_{a} \sum_{b \neq a} \frac{m_a^2 m_b}{r_{a b}^2}\left\{\frac{p_a^2}{m_a^2}+\frac{p_b^2}{m_b^2}-2 \frac{\mathbf{p}_a \cdot \mathbf{p}_b}{m_a m_b}\right\} \\
    & +\frac{1}{2} \sum_{a} \sum_{b \neq a} \sum_{c \neq a, b} \frac{m_a m_b m_c}{\left(r_{a b}+r_{b c}+r_{ac}\right)^2}\left(n_{a b}^i+n_{a c}^i\right)(n_{a b}^j+n_{c b}^j)\Big\{8 \frac{p_{ai} p_{cj}}{m_a m_c}-16 \frac{p_{aj} p_{ci}}{m_a m_c} +3 \frac{p_{ai} p_{bj}}{m_a m_b}\\
    & +4 \frac{p_{ci} p_{cj}}{m_c^2}+\frac{p_{ai} p_{aj}}{m_a^2}\Big\}  +\frac{1}{2} \sum_{a} \sum_{b \neq a} \sum_{c \neq a, b} \frac{m_a m_b m_c}{\left(r_{a b}+r_{b c}+r_{c a}\right) r_{a b}}\Bigg\{8 \frac{\mathbf{p}_a \cdot \mathbf{p}_c-\left(\mathbf{n}_{a b} \cdot \mathbf{p}_a\right)\left(\mathbf{n}_{a b} \cdot \mathbf{p}_c\right)}{m_a m_c} \\
    & -3 \frac{\mathbf{p}_a \cdot \mathbf{p}_b-\left(\mathbf{n}_{a b} \cdot \mathbf{p}_a\right)\left(\mathbf{n}_{a b} \cdot \mathbf{p}_b\right)}{m_a m_b} -4 \frac{p_c^2-\left(\mathbf{n}_{a b} \cdot \mathbf{p}_c\right)^2}{m_c^2}-\frac{p_a^2-\left(\mathbf{n}_{a b} \cdot \mathbf{p}_a\right)^2}{m_a^2}\Bigg\} \\
    &- \frac{3}{8} \sum_{a} \sum_{b \neq a} \sum_{c \neq b} \sum_{d \neq c} \frac{m_a m_b m_c m_d}{r_{ab} r_{bc} r_{cd}} - \frac{1}{4} \sum_{a} \sum_{b \neq a} \sum_{c \neq a} \sum_{d \neq a} \frac{m_a m_b m_c m_d}{r_{ab} r_{ac} r_{ad}} \\
    &-\frac{1}{4} \sum_{a} \sum_{b \neq a} \frac{m_a^2 m_b^2}{r_{a b}^3} -\frac{1}{64} \sum_{a} \sum_{b \neq a} \sum_{c \neq a, b} \frac{m_a^2 m_b m_c}{r_{a b}^3 r_{a c}^3 r_{b c}}\Big\{18 r_{a b}^2 r_{a c}^2-60 r_{a b}^2 r_{b c}^2-24 r_{a b}^2 r_{a c}\left(r_{a b}+r_{b c}\right) \\
    &+60 r_{a b} r_{a c} r_{b c}^2 +56 r_{a b}^3 r_{b c} -72 r_{a b} r_{b c}^3 +35 r_{b c}^4+6 r_{a b}^4\Big\} \\
    &-\frac{1}{64} \sum_a \sum_{b \neq a} \sum_{c \neq a,b} \sum_{d \neq a,b,c}
    \frac{m_a m_b m_c m_d}{r_{ab}^{3}r_{cd}^{3} r_{ad}}
    \Bigg\{35r_{ad}^{4}+34r_{ad}^{2}r_{bc}^{2}+10r_{ab}^{2}r_{ad}^{2}+r_{ac}^{2}r_{bd}^{2}-70r_{ad}^{2}r_{bd}^{2}-10r_{bd}^{2}r_{cd}^{2} \\
    &+17r_{ab}^{2}r_{cd}^{2} - 32 r_{ab}^2 r_{ad} r_{cd}
    +16r_{cd}\Bigg(
    \frac{
    (r_{cd}^{2}-8r_{ac}r_{ad})
    (r_{bd}^{2}-r_{ab}^{2}-r_{ac}r_{ad})
    }{
    r_{ac}+r_{ad}+r_{cd}
    } - \frac{
    r_{ab}^2r_{ac}(r_{ab}r_{cd}+4r_{ad}r_{bc})
    }{
    r_{ab}r_{cd}+r_{ac}r_{bd}+r_{ad}r_{bc}
    }
    \Bigg)\Bigg\} \\
    &- \sum_a \sum_{b \neq a} \sum_{c \neq a,b} \sum_{d \neq a,b,c}
    \frac{m_a m_b m_c m_d}{r_{ab}^{3}r_{cd}^{3}}\Delta_{\{a,b,c,d\}}\Phi_{\{a,b,c,d\}}.
\end{aligned}
\label{eq:2PN_hamiltonian}
\end{equation}
We also correct an error in the term $2(\mathbf{n}_{ab}\cdot\mathbf{p}_b) (\mathbf{n}_{ac}\cdot\mathbf{p}_c)/(m_a m_c)$, replacing $m_a m_c$ by $m_b m_c$. This error is present in Refs.~\cite{LoustoNakano2008, GalavizBruegmann2011, Bonetti2016, HeinzeSchaeferBruegmann2026} but is absent from the earlier expressions in Refs.~\cite{OhtaEtAl1974b, OhtaEtAl1974a, Schaefer1987}.

\end{widetext}

\begin{figure}[htp!]
    \centering
    \includegraphics[width=1.0\linewidth]{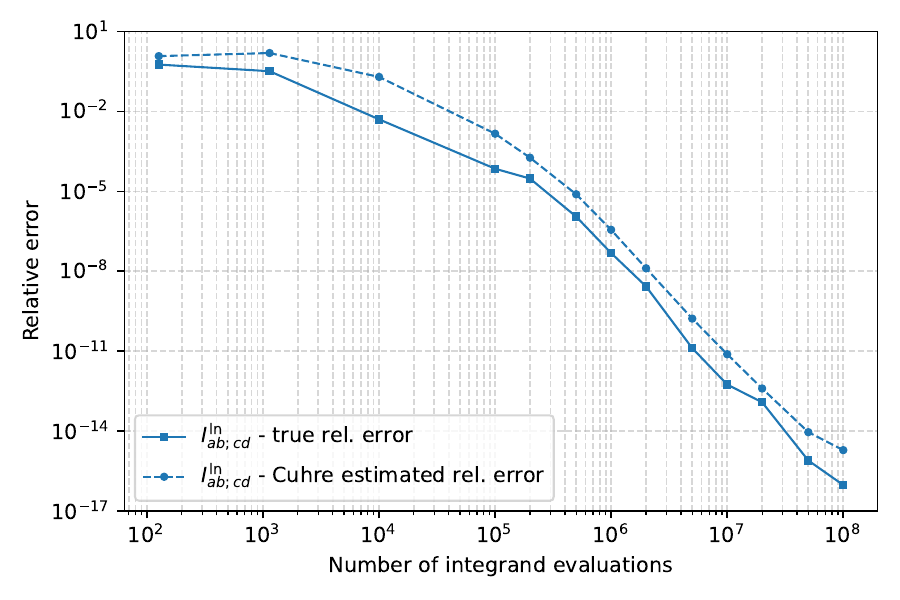}
    \caption{The true and estimated relative errors for the numerical evaluation of $I^{\mathrm{ln}}_{ab;cd}$ compared to the result of Eq.~\eqref{eq:I_ln_solution} for a generic random four-particle configuration.}
    \label{fig:integral_accuracy}
\end{figure}
\section{Numerical validation}
\label{sec:numerical_validation}
We validated the analytic expressions with independent numerical comparisons. For a generic four-particle configuration, we evaluated the spatial integral representation in Eq.~\eqref{eq:ln_integral} using the \textsc{Cuhre} algorithm of the \textsc{CUBA} library \cite{Hahn2005}, following the approach of Ref.~\cite{HeinzeSchaeferBruegmann2026}. The comparison with the analytic result of Eq.~\eqref{eq:I_ln_solution} for different numbers of integrand evaluations is shown in Fig.~\ref{fig:integral_accuracy}. Both the numerical integrals and the analytic expressions were evaluated in quadruple precision.

We additionally tested 1000 randomly generated noncoincident four-particle configurations, divided equally among generic, hierarchical, coplanar, and collinear configurations. Eq.~\eqref{eq:H_ln_compact} was compared with the logarithmic contribution to Eq.~\eqref{eq:UTT4}, evaluated numerically in quadruple precision using the two-dimensional parameter integral in Eq.~\eqref{eq:2d_parameter_integral}. Adding the explicit nonlogarithmic terms then provided the comparison with the complete expression in Eq.~\eqref{eq:UTT4_compact}. The parameter integrals were evaluated using tensor-product Gauss--Legendre quadrature with automatic order refinement. Convergence was assessed by comparing successive quadrature orders, with a target estimated relative error of $10^{-22}$ for the numerical evaluations used to test Eqs.~\eqref{eq:H_ln_compact} and \eqref{eq:UTT4_compact}. All configurations satisfied the convergence criteria, and the highest order used was 292 per integration variable. Compensated summation was used to reduce roundoff. With this, we obtained mean relative deviations of $3.43\times10^{-28}$ and $1.67\times10^{-26}$ for Eqs.~\eqref{eq:H_ln_compact} and \eqref{eq:UTT4_compact}, respectively. The corresponding maximum relative deviations were $3.71\times10^{-26}$ and $4.93\times10^{-24}$. Evaluations at 100-digit precision showed that, besides numerical errors in the integral evaluation, roundoff in the analytic evaluations also contributes significantly to the residual differences.

\section{Concluding remarks}
\label{sec:conclusion}
We have obtained a closed-form expression for the remaining integral in the general $N$-body 2PN Hamiltonian in the ADM gauge. This completes the four-body contribution to the static potential and, together with the previously known terms, yields the complete conservative Hamiltonian for an arbitrary number of nonspinning point particles up to 2PN order. This Hamiltonian contains no unevaluated spatial integrals and provides a basis for explicitly computing the canonical equations of motion. In particular, the four-body contribution can now be evaluated and differentiated without the need for numerical quadrature. This removes a difficulty in implementing the complete 2PN dynamics in numerical studies of systems containing four or more bodies. The numerical methods of Ref.~\cite{HeinzeSchaeferBruegmann2026} certainly remain useful for independently checking the analytic expressions, but in addition may also provide a means of treating integrals and integrating the equations of motion at higher PN orders if closed-form results are unavailable.

A natural next objective is the complete conservative $N$-body Hamiltonian at 3PN order. The dissipative radiation-reaction contributions at 2.5PN and 3.5PN orders are already available for general $N$-body systems \cite{JaranowskiSchaefer1997}. Although partial results beyond the two-body problem have also been obtained at 3PN order, including $G^2 v^4$ contributions to the three-body interaction~\cite{LoebbertEtAl2021}, a complete general $N$-body treatment requires the remaining conservative interactions at that order.

\section*{Acknowledgments}
FMH has been supported by the Deutsche Forschungsgemeinschaft (DFG) under Grant No. 406116891 within the Research Training Group RTG 2522/1. FMH also acknowledges support from the German Academic Exchange Service (DAAD) through a doctoral research scholarship (``Forschungsstipendium f\"ur Doktorandinnen und Doktoranden'').

This work has made substantive use of the large language model \textsc{GPT-6 Astra} by OpenAI to assist in developing the derivations and checking the calculations presented in this work. All results were independently verified by the authors, who take full responsibility for the accuracy and content of this paper.

\section*{Data Availability}
The numerical data and code used to validate the analytic expressions presented in this work are available from the corresponding author upon reasonable request.

\bibliography{apssamp} 

@article{FutamaseItoh2007,
  author        = {Futamase, Toshifumi and Itoh, Yousuke},
  title         = {The post-{Newtonian} approximation for relativistic compact binaries},
  journal       = {Living Rev. Relativ.},
  volume        = {10},
  pages         = {2},
  year          = {2007},
  doi           = {10.12942/lrr-2007-2}
}

@article{Blanchet2024,
  author        = {Blanchet, Luc},
  title         = {Post-{Newtonian} theory for gravitational waves},
  journal       = {Living Rev. Relativ.},
  volume        = {27},
  pages         = {4},
  year          = {2024},
  doi           = {10.1007/s41114-024-00050-z},
  eprint        = {1310.1528},
  archivePrefix = {arXiv}
}

@article{GalavizBruegmann2011,
  author        = {Galaviz, Pablo and Br{\"u}gmann, Bernd},
  title         = {Characterization of the gravitational wave emission of three black holes},
  journal       = {Phys. Rev. D},
  volume        = {83},
  pages         = {084013},
  year          = {2011},
  doi           = {10.1103/PhysRevD.83.084013},
  eprint        = {1012.4423},
  archivePrefix = {arXiv}
}

@article{Will2014,
  author        = {Will, Clifford M.},
  title         = {Incorporating post-{Newtonian} effects in {$N$}-body dynamics},
  journal       = {Phys. Rev. D},
  volume        = {89},
  pages         = {044043},
  year          = {2014},
  doi           = {10.1103/PhysRevD.89.044043},
  eprint        = {1312.1289},
  archivePrefix = {arXiv},
  note          = {Erratum: \href{https://doi.org/10.1103/PhysRevD.91.029902}{Phys. Rev. D \textbf{91}, 029902 (2015)}}
}

@incollection{ADM1962,
  author        = {Arnowitt, Richard and Deser, Stanley and Misner, Charles W.},
  title         = {The dynamics of general relativity},
  editor        = {Witten, Louis},
  booktitle     = {Gravitation: An Introduction to Current Research},
  publisher     = {Wiley},
  address       = {New York},
  pages         = {227--265},
  year          = {1962},
  eprint        = {gr-qc/0405109},
  archivePrefix = {arXiv},
  note          = {Reprinted in \href{https://doi.org/10.1007/s10714-008-0661-1}{Gen. Relativ. Gravit. \textbf{40}, 1997--2027 (2008)}}
}

@article{SchaeferJaranowski2024,
  author        = {Sch{\"a}fer, Gerhard and Jaranowski, Piotr},
  title         = {{Hamiltonian} formulation of general relativity and post-{Newtonian} dynamics of compact binaries},
  journal       = {Living Rev. Relativ.},
  volume        = {27},
  pages         = {2},
  year          = {2024},
  doi           = {10.1007/s41114-024-00048-7},
  eprint        = {1805.07240},
  archivePrefix = {arXiv}
}

@article{EIH1938,
  author        = {Einstein, Albert and Infeld, Leopold and Hoffmann, Banesh},
  title         = {The gravitational equations and the problem of motion},
  journal       = {Ann. Math.},
  volume        = {39},
  pages         = {65--100},
  year          = {1938},
  doi           = {10.2307/1968714}
}

@article{OhtaEtAl1974a,
  author        = {Ohta, Tadayuki and Okamura, Hiroshi and Kimura, Toshiei and Hiida, Kichiro},
  title         = {Higher order gravitational potential for many-body system},
  journal       = {Prog. Theor. Phys.},
  volume        = {51},
  pages         = {1220--1238},
  year          = {1974},
  doi           = {10.1143/PTP.51.1220}
}

@article{OhtaEtAl1974b,
  author        = {Ohta, Tadayuki and Okamura, Hiroshi and Kimura, Toshiei and Hiida, Kichiro},
  title         = {Coordinate condition and higher order gravitational potential in canonical formalism},
  journal       = {Prog. Theor. Phys.},
  volume        = {51},
  pages         = {1598--1612},
  year          = {1974},
  doi           = {10.1143/PTP.51.1598}
}

@article{DamourSchaefer1985,
  author        = {Damour, Thibault and Sch{\"a}fer, Gerhard},
  title         = {{Lagrangians} for {$n$} point masses at the second post-{Newtonian} approximation of general relativity},
  journal       = {Gen. Relativ. Gravit.},
  volume        = {17},
  pages         = {879--905},
  year          = {1985},
  doi           = {10.1007/BF00773685}
}

@article{Schaefer1987,
  author        = {Sch{\"a}fer, Gerhard},
  title         = {Three-body {Hamiltonian} in general relativity},
  journal       = {Phys. Lett. A},
  volume        = {123},
  pages         = {336--339},
  year          = {1987},
  doi           = {10.1016/0375-9601(87)90389-6}
}

@article{LoustoNakano2008,
  author        = {Lousto, Carlos O. and Nakano, Hiroyuki},
  title         = {Three-body equations of motion in successive post-{Newtonian} approximations},
  journal       = {Class. Quantum Grav.},
  volume        = {25},
  pages         = {195019},
  year          = {2008},
  doi           = {10.1088/0264-9381/25/19/195019},
  eprint        = {0710.5542},
  archivePrefix = {arXiv}
}

@article{Chu2009,
  author        = {Chu, Yi-Zen},
  title         = {{$n$}-body problem in general relativity up to the second post-{Newtonian} order from perturbative field theory},
  journal       = {Phys. Rev. D},
  volume        = {79},
  pages         = {044031},
  year          = {2009},
  doi           = {10.1103/PhysRevD.79.044031},
  eprint        = {0812.0012},
  archivePrefix = {arXiv}
}

@article{HeinzeSchaeferBruegmann2026,
  author        = {Heinze, Felix M. and Sch{\"a}fer, Gerhard and Br{\"u}gmann, Bernd},
  title         = {{$N$}-body {2PN} {Hamiltonian} and numerical integration of the equations of motion},
  journal       = {Phys. Rev. D},
  volume        = {113},
  pages         = {104066},
  year          = {2026},
  doi           = {10.1103/9vhx-6wkk},
  eprint        = {2602.06961},
  archivePrefix = {arXiv}
}

@article{ParkEtAl2021,
  author        = {Park, Ryan S. and Folkner, William M. and Williams, James G. and Boggs, Dale H.},
  title         = {The {JPL} planetary and lunar ephemerides {DE440} and {DE441}},
  journal       = {Astron. J.},
  volume        = {161},
  pages         = {105},
  year          = {2021},
  doi           = {10.3847/1538-3881/abd414}
}

@article{WeisbergHuang2016,
  author        = {Weisberg, Joel M. and Huang, Yuping},
  title         = {Relativistic measurements from timing the binary pulsar {PSR B1913+16}},
  journal       = {Astrophys. J.},
  volume        = {829},
  pages         = {55},
  year          = {2016},
  doi           = {10.3847/0004-637X/829/1/55},
  eprint        = {1606.02744},
  archivePrefix = {arXiv}
}

@article{KramerEtAl2021,
  author        = {Kramer, M. and Stairs, I. H. and Manchester, R. N. and Wex, N. and Deller, A. T. and others},
  title         = {Strong-field gravity tests with the {Double Pulsar}},
  journal       = {Phys. Rev. X},
  volume        = {11},
  pages         = {041050},
  year          = {2021},
  doi           = {10.1103/PhysRevX.11.041050},
  eprint        = {2112.06795},
  archivePrefix = {arXiv}
}

@article{BlanchetDamourIyer1995,
  author        = {Blanchet, Luc and Damour, Thibault and Iyer, Bala R.},
  title         = {Gravitational waves from inspiralling compact binaries: Energy loss and waveform to second-post-{Newtonian} order},
  journal       = {Phys. Rev. D},
  volume        = {51},
  pages         = {5360--5386},
  year          = {1995},
  doi           = {10.1103/PhysRevD.51.5360},
  eprint        = {gr-qc/9501029},
  archivePrefix = {arXiv},
  note          = {Erratum: \href{https://doi.org/10.1103/PhysRevD.54.1860}{Phys. Rev. D \textbf{54}, 1860 (1996)}}
}

@article{NaozEtAl2013,
  author        = {Naoz, Smadar and Kocsis, Bence and Loeb, Abraham and Yunes, Nicol{\'a}s},
  title         = {Resonant post-{Newtonian} eccentricity excitation in hierarchical three-body systems},
  journal       = {Astrophys. J.},
  volume        = {773},
  pages         = {187},
  year          = {2013},
  doi           = {10.1088/0004-637X/773/2/187},
  eprint        = {1206.4316},
  archivePrefix = {arXiv}
}

@article{KupiEtAl2006,
  author        = {Kupi, G{\'a}bor and Amaro-Seoane, Pau and Spurzem, Rainer},
  title         = {Dynamics of compact object clusters: a post-{Newtonian} study},
  journal       = {Mon. Not. R. Astron. Soc.},
  volume        = {371},
  pages         = {L45--L49},
  year          = {2006},
  doi           = {10.1111/j.1745-3933.2006.00205.x},
  eprint        = {astro-ph/0602125},
  archivePrefix = {arXiv}
}

@article{RantalaEtAl2023,
  author        = {Rantala, Antti and Naab, Thorsten and Rizzuto, Francesco Paolo and Mannerkoski, Matias and Partmann, Christian and Lautensch{\"u}tz, Kristina},
  title         = {{BIFROST}: simulating compact subsystems in star clusters using a hierarchical fourth-order forward symplectic integrator code},
  journal       = {Mon. Not. R. Astron. Soc.},
  volume        = {522},
  pages         = {5180--5203},
  year          = {2023},
  doi           = {10.1093/mnras/stad1360},
  eprint        = {2210.02472},
  archivePrefix = {arXiv}
}

@article{SpurzemKamlah2023,
  author        = {Spurzem, Rainer and Kamlah, Albrecht},
  title         = {Computational methods for collisional stellar systems},
  journal       = {Living Rev. Comput. Astrophys.},
  volume        = {9},
  pages         = {3},
  year          = {2023},
  doi           = {10.1007/s41115-023-00018-w},
  eprint        = {2305.11606},
  archivePrefix = {arXiv}
}

@article{Levi:2018nxp,
  author        = {Levi, Mich{\`e}le},
  title         = {Effective field theories of post-{Newtonian} gravity: A comprehensive review},
  journal       = {Rep. Prog. Phys.},
  volume        = {83},
  pages         = {075901},
  year          = {2020},
  doi           = {10.1088/1361-6633/ab12bc},
  eprint        = {1807.01699},
  archivePrefix = {arXiv}
}

@article{BlanchetDamour1986,
  author        = {Blanchet, Luc and Damour, Thibault},
  title         = {Radiative gravitational fields in general relativity. {I}. General structure of the field outside the source},
  journal       = {Philos. Trans. R. Soc. Lond. A},
  volume        = {320},
  pages         = {379--430},
  year          = {1986},
  doi           = {10.1098/rsta.1986.0125}
}

@article{PoujadeBlanchet2002,
  author        = {Poujade, Olivier and Blanchet, Luc},
  title         = {Post-{Newtonian} approximation for isolated systems calculated by matched asymptotic expansions},
  journal       = {Phys. Rev. D},
  volume        = {65},
  pages         = {124020},
  year          = {2002},
  doi           = {10.1103/PhysRevD.65.124020},
  eprint        = {gr-qc/0112057},
  archivePrefix = {arXiv}
}

@article{Bonetti2016,
  author        = {Bonetti, Matteo and Haardt, Francesco and Sesana, Alberto and Barausse, Enrico},
  title         = {Post-{Newtonian} evolution of massive black hole triplets in galactic nuclei -- {I}. Numerical implementation and tests},
  journal       = {Mon. Not. R. Astron. Soc.},
  volume        = {461},
  pages         = {4419--4434},
  year          = {2016},
  doi           = {10.1093/mnras/stw1590},
  eprint        = {1604.08770},
  archivePrefix = {arXiv}
}

@article{LoebbertEtAl2021,
  author        = {Loebbert, Florian and Plefka, Jan and Shi, Canxin and Wang, Tianheng},
  title         = {Three-body effective potential in general relativity at second post-{Minkowskian} order and resulting post-{Newtonian} contributions},
  journal       = {Phys. Rev. D},
  volume        = {103},
  pages         = {064010},
  year          = {2021},
  doi           = {10.1103/PhysRevD.103.064010},
  eprint        = {2012.14224},
  archivePrefix = {arXiv}
}

@article{JaranowskiSchaefer1997,
  author        = {Jaranowski, Piotr and Sch{\"a}fer, Gerhard},
  title         = {Radiative 3.5 post-{Newtonian} {ADM} {Hamiltonian} for many-body point-mass systems},
  journal       = {Phys. Rev. D},
  volume        = {55},
  pages         = {4712--4722},
  year          = {1997},
  doi           = {10.1103/PhysRevD.55.4712}
}

@article{PatiWill2000,
  author        = {Pati, Michael E. and Will, Clifford M.},
  title         = {Post-{Newtonian} gravitational radiation and equations of motion via direct integration of the relaxed {Einstein} equations: Foundations},
  journal       = {Phys. Rev. D},
  volume        = {62},
  pages         = {124015},
  year          = {2000},
  doi           = {10.1103/PhysRevD.62.124015},
  eprint        = {gr-qc/0007087},
  archivePrefix = {arXiv}
}

@article{HuangYangNi2026,
  author        = {Huang, Hongkun and Yang, Jie and Ni, Wei-Tou},
  title         = {{2PN} point-mass {$N$}-body equations of motion in harmonic gauge: A computable formulation},
  journal       = {Phys. Rev. D},
  year          = {2026},
  doi           = {10.1103/y8nj-tmz3},
  eprint        = {2608.20193},
  archivePrefix = {arXiv},
  note          = {Accepted for publication}
}

@article{Hahn2005,
  author        = {Hahn, Thomas},
  title         = {{Cuba}---a library for multidimensional numerical integration},
  journal       = {Comput. Phys. Commun.},
  volume        = {168},
  pages         = {78--95},
  year          = {2005},
  doi           = {10.1016/j.cpc.2005.01.010},
  eprint        = {hep-ph/0404043},
  archivePrefix = {arXiv}
}

@misc{SupplementalMaterial,
  note          = {See the Supplemental Material provided as an ancillary file for the derivation of Eq.~(12).}
}

@article{JaranowskiSchafer1998,
  author        = {Jaranowski, Piotr and Sch{\"a}fer, Gerhard},
  title         = {Third post-{Newtonian} higher order {ADM} {Hamilton} dynamics for two-body point-mass systems},
  journal       = {Phys. Rev. D},
  volume        = {57},
  pages         = {7274--7291},
  year          = {1998},
  doi           = {10.1103/PhysRevD.57.7274},
  eprint        = {gr-qc/9712075},
  archivePrefix = {arXiv},
  note          = {Erratum: \href{https://doi.org/10.1103/PhysRevD.63.029902}{Phys. Rev. D \textbf{63}, 029902 (2001)}}
}

\end{document}